\makeatletter
\def\@copyrightspace{\relax}
\makeatother

\documentclass{sigchi-ext}
\usepackage[T1]{fontenc}
\usepackage{textcomp}
\usepackage[scaled=.92]{helvet} 
\usepackage{graphicx} 
\usepackage{balance}  
\usepackage{booktabs} 
\usepackage{ccicons}  
\usepackage{ragged2e} 
\usepackage{hyperref}

\def\plaintitle{BlackBox Toolkit: Intelligent Assistance to UI Design}

\def\emptyauthor{}
\def\plainkeywords{UI Element Dataset, Neural Networks, Deep Learning, Sketch Detection, Sketch Recognition, Blueprints, UI Layout, Prototyping}

\title{BlackBox Toolkit: Intelligent Assistance to UI Design}

\numberofauthors{2}
\author{%
  \alignauthor{%
    \textbf{Vinoth Pandian Sermuga Pandian}\\
    \affaddr{Fraunhofer FIT - UCC} \\
    \affaddr{Sankt Augustin, Germany} \\
    \email{pandian@fit.fraunhofer.de} }  \vfil    \alignauthor{%
    \textbf{Sarah Suleri}\\
    \affaddr{Fraunhofer FIT - UCC} \\
    \affaddr{Sankt Augustin, Germany} \\
    \email{suleri@fit.fraunhofer.de} 
  }
}

\definecolor{linkColor}{RGB}{6,125,233}
\hypersetup{%
  pdftitle={\plaintitle},
  pdfauthor={\emptyauthor},
  pdfkeywords={\plainkeywords},
  bookmarksnumbered,
  pdfstartview={FitH},
  colorlinks,
  citecolor=black,
  filecolor=black,
  linkcolor=black,
  urlcolor=linkColor,
  breaklinks=true,
}

\begin{document}

\CopyrightYear{2020}
\setcopyright{rightsretained}
\conferenceinfo{CHI'20, Workshop on Artificial Intelligence for HCI: A Modern Approach}{April  25--30, 2020, Honolulu, HI, USA}
\isbn{}
\doi{https://sites.google.com/view/ai4hci/accepted-papers}
\copyrightinfo{\acmcopyright}

\maketitle

\RaggedRight{}

\begin{abstract}
    User Interface (UI) design is an creative process that involves considerable reiteration and rework. Designers go through multiple iterations of different prototyping fidelities to create a UI design. In this research, we propose to modify the UI design process by assisting it with artificial intelligence (AI). We propose to enable AI to perform repetitive tasks for the designer while allowing the designer to take command of the creative process. This approach makes the machine act as a black box that intelligently assists the designers in creating UI design. We believe this approach would greatly benefit designers in co-creating design solutions with AI.
\end{abstract}




\section{Introduction}
The user interface (UI) acts as the bridge between a human and a machine. \marginpar{ 
\begin{quote}
    "Just as the Industrial Revolution freed up a lot of humanity from physical drudgery I think AI has the potential to free up humanity from a lot of the mental drudgery." \\ --\textit{Andrew Ng}
\end{quote} 
}
It acts as a translator mediating between two worlds: one disorderly, irrational, but adept at noticing patterns; another structured, analytical, however inept in pattern-finding (as of now). A UI designer is an architect who designs this bridge between man and machine.

The most important job of a UI designer is not to find a balance between both worlds; instead, reduce the mental load of the one with issues and emotions and try to fit it into the confinements and restrictions placed by the other. This task of designing such interfaces is strenuous. Among the numerous ways of creating user interfaces to satisfy both the worlds involved, the most commonly used technique is user-centered design.

In user-centered design, users are kept at the heart of design, and designers attempt to satisfy their needs by analyzing the usage context, user needs, and requirements before starting the design process. Then during the design process, designers go through multiple fidelities of prototypes. Starting from low-fidelity (lo-fi) freehand sketches to medium-fidelity (me-fi) digital images and finally to high-fidelity (hi-fi) interactive screens or code. The different fidelities have their strengths and weaknesses. For example, lo-fi is cheap and supports ideating different designs quickly; however, it does not do justice to the final look-and-feel of the system. Similarly, me-fi contains the most necessary information and is quicker to create than hi-fi; however, it is hard to create multiple design variations compared to lo-fi. Hi-fi resembles the final product, but the workload of creating such a system is humongous, and it is hard to create multiple design modifications to test the system.

Several researchers attempt to solve the issues in this design process. After the advancements of AI, one solution is to automate this whole process - the designers sketch a UI, and then the machine analyses the sketch and generates the hi-fi code. This interesting approach, however, has one major flaw. The whole system acts like a transformer to the designer who enters their sketch and gets a corresponding code with no control over tweaking the intricate design details. As a solution, we approach the same problem domain and propose two solutions. Our goal is to enable the machine to perform repetitive tasks for the designer while allowing the designer to take command of the creative process. This approach makes the machine act as a black box that intelligently assists the designers in UI design. The machine's role in our proposed design process is no different from the apprentices of renaissance art maestros. The apprentice's task to assist the artist in preparing materials and executing the less critical and quite tedious decorative parts of frescoes or statues. We believe this approach would greatly benefit designers where AI and human co-create creative solutions.

\section{Research Focus}
Our focus in this research is on how to move designers as the drivers of creativity and let AI assist designers in the UI design process. Our primary research question in this research is, "How can we automate the UI design process while allowing UI designers to control the design details." We address this question by using artificial intelligence to automate the transformation of different fidelities of UI design. In the following sections, we expand on each of our proposed solutions, the challenges we faced with implementation and the benefit of that solution.

\section{MetaMorph}

MetaMorph is a UI element detector, created with a Deep Neural Network (DNN) object detection model \cite{eve}. MetaMorph detects constituent UI element categories and their position from a lo-fi sketch using a fine-tuned RetinaNet object detection model trained with a dataset of 125,000 synthetically generated lo-fi sketches.

\subsection{Challenge \& Solution}
The major challenge in creating MetaMorph was the dataset. We required a large scale lo-fi sketch dataset, which was non-existent. Therefore, we collected UI element sketches from 350 participants using paper and digital questionnaires. Then we processed this data and labeled it to create the UISketch dataset\footnote{\url{https://www.kaggle.com/vinothpandian/uisketch}} \cite{syn}. This dataset contains 5,917 sketches of 19 UI elements. However, this labeled dataset is only useful for classifying UI elements; but, a UI element detector would need ground truth of both the identity and location of UI elements in a lo-fi sketch.

\begin{marginfigure}
    \includegraphics[width=\marginparwidth]{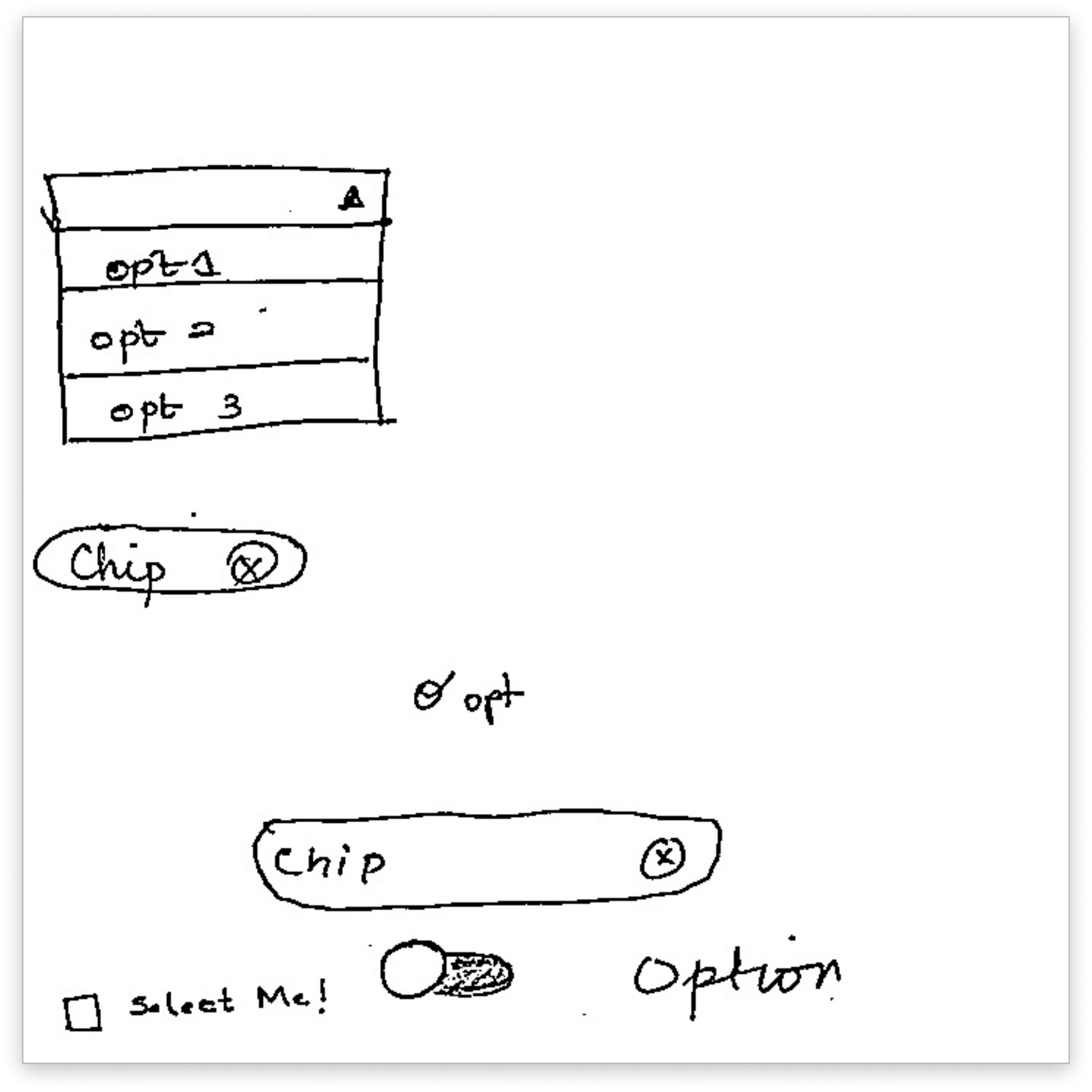}
    \caption{Sample generated synthetic data}
    \label{fig:syn}
\end{marginfigure}

As a solution, we created Syn\footnote{\url{https://www.kaggle.com/vinothpandian/syn-dataset}}, a large-scale synthetic dataset containing 125,000 synthetically generated lo-fi sketches \cite{syn}. To create Syn, we randomly chose UI elements from the labeled UISketch dataset and stitched them in random locations with random scaling (Figure \ref{fig:syn}). This random allocation of elements in an image is similar to the pre-processing and data augmentation techniques used in improving detection metrics of object detection models. We used Syn to train the MetaMorph UI element detector.

We then collected 200 lo-fi sketches to evaluate MetaMorph. The evaluation results indicate that MetaMorph detects UI elements from lo-fi sketches with 63.5\% mAP. MetaMorph\footnote{\url{https://metamorph.designwitheve.com/}} is available as an open web API\footnote{\url{http://api.metamorph.designwitheve.com/}}. We have also open-sourced the UISketch dataset and Syn.

\begin{marginfigure}
    \includegraphics[width=\marginparwidth]{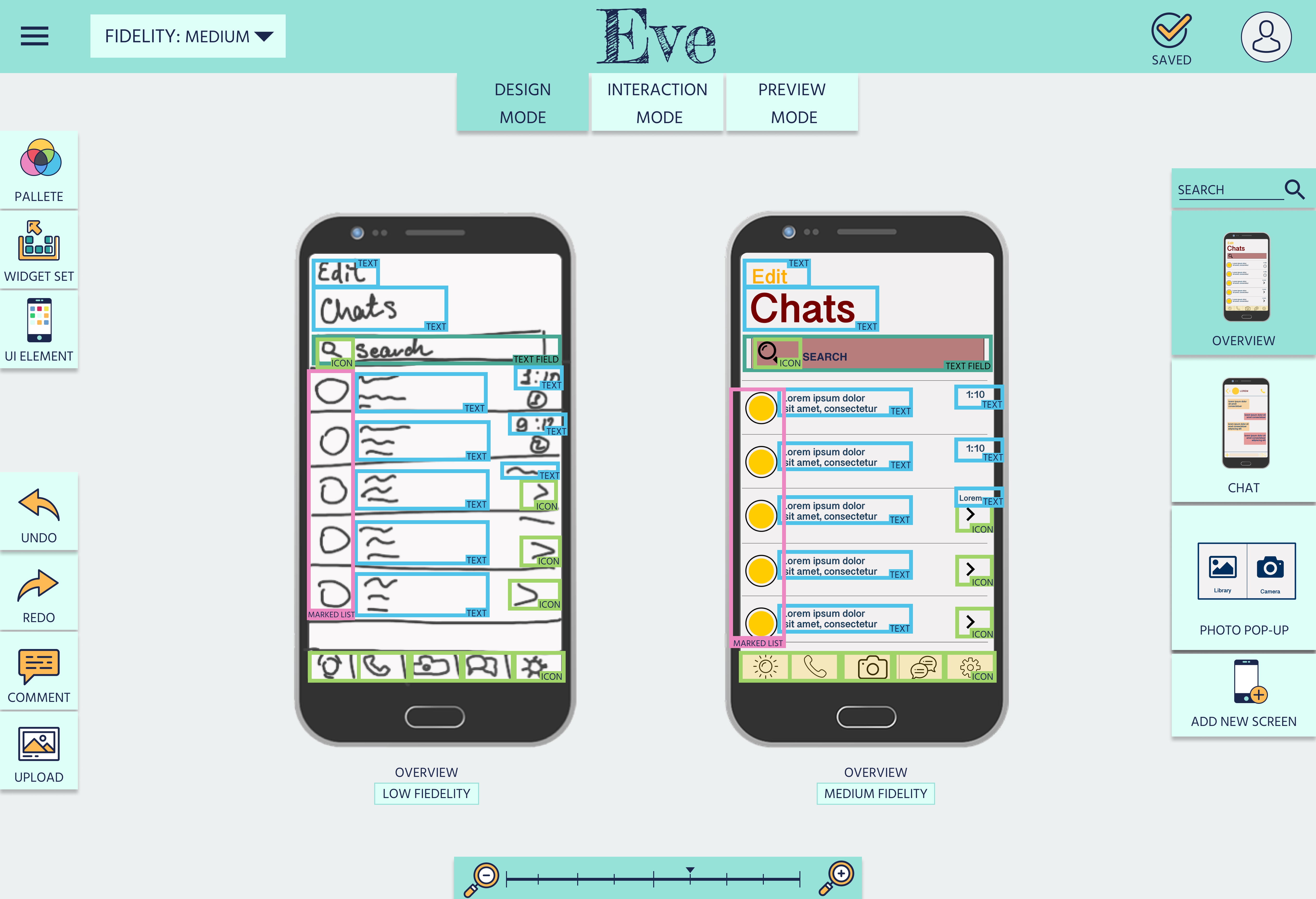}
    \caption{Usage of MetaMorph to detect lo-fi elements and transforming them to me-fi in Eve}
    \label{LoFi to MeFi}
\end{marginfigure}

\subsection{Benefits}
By detecting the UI elements present in a lo-fi sketch, the lo-fi prototype can be converted to me-fi or hi-fi instead of fully automating the process (Figure \ref{LoFi to MeFi}). MetaMorph enabled us to create Eve, a prototyping workbench \cite{eve} where a designer can sketch or upload a lo-fi, which will be converted to me-fi and later to hi-fi by means of UI element detection. Eve enables the designer to control the styling of the UI in me-fi and progress it to hi-fi android XML code.

\section{Blu}
Blu is a tool that generates UI layout information from UI screenshots. With the detected information, it enables conversion of UI screenshots to blueprints and editable vector graphics \cite{blu}. Herring et al. demonstrate the benefits and role of design examples in different aspects of the UI design process \cite{getting_inspired}.  In this research, we expand on this idea, and from a UI design example screenshot, we detect UI element categories, positions, grouping information, and layout using deep neural networks.

\subsection{Challenge \& Solution}
We faced two significant challenges in implementing Blu: dataset and layout detection.

Fortunately, as a solution to dataset issue, Deka et al. collected a large scale android UI screenshot dataset, RICO \cite{rico}. RICO contains 72k UI screenshots with UI element hierarchy and semantic annotations. However, RICO was annotated using an automated approach; therefore, the annotations are sometimes mislabelled. If RICO is directly used as training data of a DNN, these mistakes propagate and provide inadequate results. Therefore, we had to re-annotate the elements and layout information for a subset of RICO to train Blu.

Another challenge in creating Blu is identifying the layout information. A UI designer by education and practice groups and aligns UI elements while creating a me-fi prototype. They group UI elements mostly based on gestalt laws. This layout information further helps the front-end developers to create the hi-fi with the constraints placed on them by programming languages and development environments. However, there is no algorithmic way to automate the layout of UI elements using gestalt law yet.  To solve this issue, we are attempting to automate the alignment process algorithmically based on gestalt laws.

\begin{marginfigure}

    \vspace{145pt}
    \includegraphics[trim = 30 40 30 20,
        clip, width=\marginparwidth]{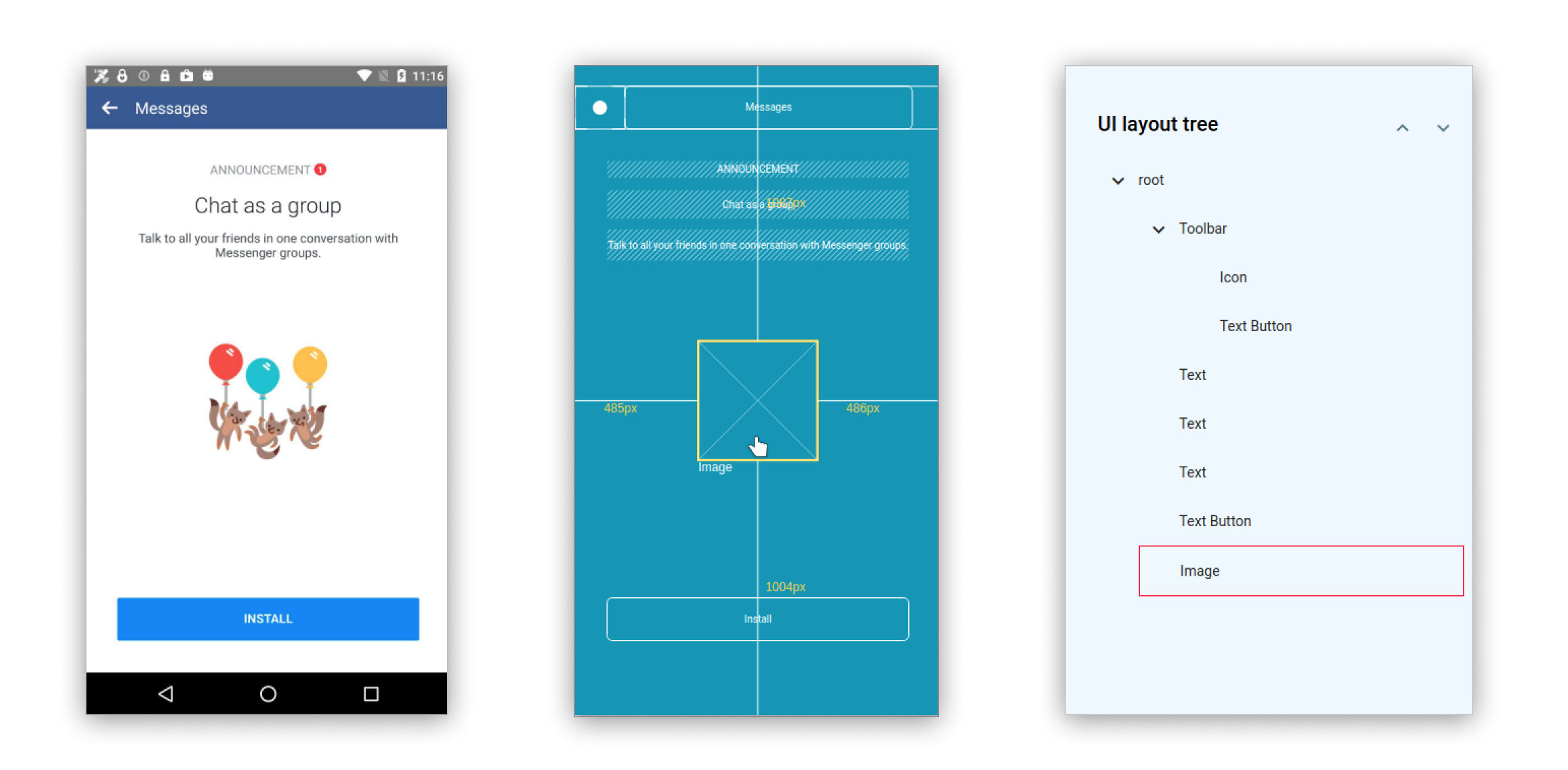}
    \caption{UI screen (left) and its respective blueprint (middle) and UI layout tree (right) created using Blu.}
    \label{fig:teaser}
\end{marginfigure}

\subsection{Benefits}
Blu reduces the rework of UI designers for starting a design from scratch \cite{blu}. It also assists designers to generate blueprints of UI design and convey the design information to developers. To demonstrate Blu, we created a web application\footnote{\url{https://blu.blackbox-toolkit.com/}} that utilizes the annotations from RICO and generates a blueprint (Figure \ref{fig:teaser}). This web app helps to convey the design and layout information of UI screen.

\section{Summary \& Future work}
In this paper, we presented our research on utilizing AI to assist designers in the UI designing process. We introduce the first two of our tools (MetaMorph and Blu) from our proposed solution, BlackBox toolkit. This research is an exploration of applying bleeding-edge AI research in human-computer interaction domain. This ongoing research is at its incipient phase with two tools. Further, we are planning to ideate and implement similar tools, such as generating UIs and automatic detection of accessibility issues in UIs.

By this research, we are looking forward to a future where humans and AI collaborate in creative tasks similar to the analytical tasks.


\bibliographystyle{SIGCHI-Reference-Format}
\bibliography{references}

\end{document}